\documentclass[10pt,twocolumn,letterpaper]{article}

\usepackage[pagenumbers]{cvpr} 
\usepackage{times}
\usepackage{epsfig}
\usepackage{amsmath}
\usepackage{amssymb}
\usepackage{sidecap}

\usepackage[utf8]{inputenc}
\usepackage{graphicx}
\usepackage{amsfonts}
\usepackage{bbm}
\usepackage{amsfonts}
\usepackage{bm} 
\usepackage{booktabs} 
\usepackage{multirow}
\usepackage{rotating}
\usepackage{pifont}
\usepackage{wrapfig}
\usepackage{tikz,pgfplots,pgfplotstable}
\usepackage{caption}

\definecolor{darkgreen}{rgb}{0.0, 0.7, 0.0}
\definecolor{col1}{rgb}{0.04314, 0.51765, 0.64706}
\definecolor{col2}{rgb}{0.96471, 0.78431, 0.37255}
\definecolor{col3}{rgb}{0.43529, 0.30588, 0.48627}
\definecolor{col4}{rgb}{0.61569, 0.84706, 0.4}
\definecolor{col5}{rgb}{0.79216, 0.27843, 0.18431}
\definecolor{col6}{rgb}{1., 0.62745, 0.33725}
\definecolor{col7}{rgb}{0.55294, 0.86667, 0.81569}

\newcommand{\cmark}{\color{darkgreen} \ding{51}}%
\newcommand{\xmark}{\color{red} \ding{55}}%

\newcommand{\shortcite}[1]{\cite{#1}}

\def\X{\mathbf X}
\def\ind{\delta_0}

\usepackage[pagebackref,breaklinks,colorlinks,bookmarks=false]{hyperref}

\usepackage[capitalize]{cleveref}
\crefname{section}{Sec.}{Secs.}
\Crefname{section}{Section}{Sections}
\Crefname{table}{Table}{Tables}
\crefname{table}{Tab.}{Tabs.}

\def\cvprPaperID{4187} 
\def\confName{CVPR}
\def\confYear{2024}

\begin{document}

\title{PhysGraph: Physics-Based Cloth Enhancement Using Graph Neural Networks}


\author{Oshri Halimi\\
 Technion – Israel Institute of Technology \\ and Meta Reality Labs Research\\
 \and
 Egor Larionov\\
 Meta Reality Labs Research\\
 \and
 Zohar Barzelay\\
 Meta Reality Labs Research\\
 \and
 Philipp Herholz\\
 Meta Reality Labs Research\\
 \and
 Tuur Stuyck\\
 Meta Reality Labs Research
}



\twocolumn[{%
\renewcommand\twocolumn[1][]{#1}%
\maketitle
\vspace{-0.5cm}
\centering
    \includegraphics[width=0.92\textwidth]{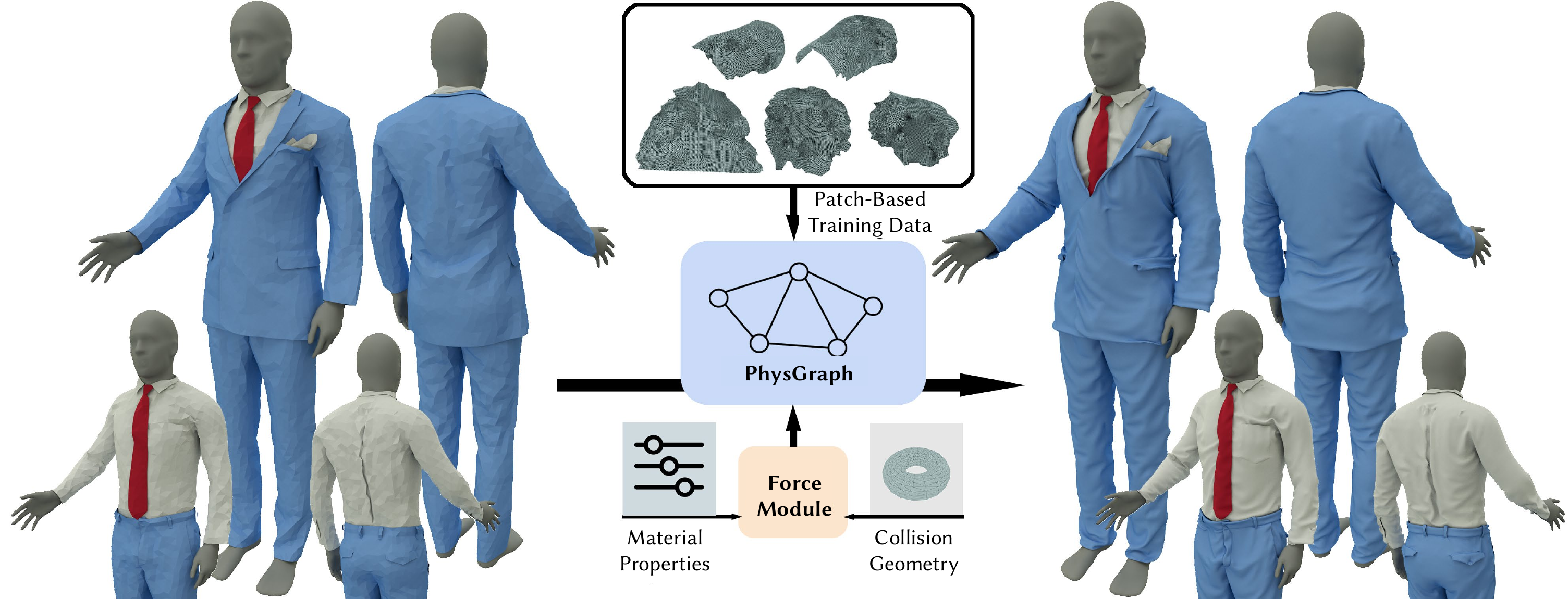}
    \captionof{figure}{\emph{PhysGraph} leverages a recurrent graph neural network in order to minimize user-provided energy potentials in a topology- and force-independent way. We demonstrate our method on enhancing coarse resolution cloth geometry with physics-based high resolution detail. The figure shows the coarse geometry on the left and the enhanced output on the right. Our method is capable of handling body and self collisions as well as the elastic potentials of cloth. By explicitly separating the force modeling and the integration process, we obtain high generalization power. The method is trained on local patches with only forces resulting from the garment elastic potential. We show several examples to demonstrate that our method allows to modify material properties and collision geometry during inference, all leveraging a single model trained on only a limited number of potentials.}
    \vspace*{0.3cm}
    \label{fig:teaser}
}]

\begin{abstract}
Physics-based simulation of mesh based domains remains a challenging task. State-of-the-art techniques can produce realistic results but require expert knowledge. A major bottleneck in many approaches is the step of integrating a potential energy in order to compute velocities or displacements. Recently, learning based method for physics-based simulation have sparked interest with graph based approaches being a promising research direction. One of the challenges for these methods is to generate models that are mesh independent and generalize to different material properties. Moreover, the model should also be able to react to unforeseen external forces like ubiquitous collisions. Our contribution is based on a simple observation: evaluating forces is computationally relatively cheap for traditional simulation methods and can be computed in parallel in contrast to their integration. If we learn how a system reacts to forces \emph{in general}, irrespective of their origin, we can learn an efficient \emph{integrator} that can predict state changes due to the total forces with high generalization power. We effectively factor out the physical model behind resulting forces by relying on an opaque \emph{force module}. We demonstrate that this idea leads to a learnable module that can be trained on basic internal forces of small mesh patches and, at inference time, generalizes to different mesh topologies, resolutions, material parameters and unseen forces like collisions. We focus our exposition on the detail enhancement of coarse clothing geometry which has many applications including computer games, virtual reality and virtual try-on. 
\end{abstract}

\section{Introduction}
Physics-based simulation has made significant advances in the last decades and it is now possible to re-create highly realistic physical phenomena using computer models. Current simulation techniques provide us with a method that generalizes to novel settings. However, these simulations are difficult to compute and can require extensive manual interventions in order to obtain the desired results. On the other hand, with the proliferation of machine learning and data-driven techniques, there has been an increased interest in recreating physical phenomena using neural networks \cite{sanchez2020learning, allen2022physical, allen2022graph, pfaff2020learning}. However, it is hard to generalize these neural models to account for all the variety that can occur as it might not be fully covered in the dataset or overfitting might occur. 
This motivates us to leverage the strengths of both approaches to design a neural model that uses physics-based information to produce a generalizable and widely applicable method. In this paper, we present a solution that improves on research in this direction.

We focus on the specific application of modeling garment deformations. The ability to model garments is crucial for telepresence, games, virtual try-on and other applications. Cloth state prediction using data-driven approaches is a long standing and notoriously difficult problem, due to the high variability in garment shape, deformation, and discontinuities caused by frequent collisions against the body and within the cloth itself. Additionally, it remains cumbersome to obtain the required training data \cite{halimi2022garment}. Despite this, many advances have been made \cite{santesteban2021self, vidaurre2020fully, bertiche2021pbns, bertiche2022neural, santesteban2022snug, santesteban2019learning}, which allow us to produce garment configurations based on the skeleton pose and body shape as input. Unfortunately these methods rely on networks trained on specific garments, and thus do not generalize well. Oftentimes, they are limited to modelling tight fitting clothing as it leverages a skinning model with respect to the body skeleton. To address these limitations, others have presented approaches for the animation of loose clothing with neural networks \cite{zhang2021dynamic}, or by leveraging real-time physics-based cloth simulation \cite{stuyck2018cloth} with a learned neural rendering pass to obtain realistic looking clothing that generalizes to new motion and body shapes \cite{xiang}. 
Despite recent progress, many limitations still remain. Methods are often limited to fixed underlying body skeletons, fixed topologies, and material properties and do not handle collisions gracefully. 

In an effort to obtain better generalization, these observations motivate us to explore the potential of combining machine learning and physics-based techniques further and exploiting knowledge about the physical system directly, instead of learning the relationship implicitly. The core idea of our proposed method is to factor out the force specific components from the integration process. Force computations are computationally relatively more efficient for physics-based simulation methods and can be computed in parallel in contrast to their integration. Leveraging this design, we can then learn the integration procedure. This approach fundamentally prevents overfitting and has the ability to generalize to novel forces. We implement this design by using a message passing graph neural network \cite{sanchez2020learning} that can integrate physics-based forces provided by an external force module on arbitrary topologies. The approach is agnostic to the specific material models and we demonstrate that it generalizes to unseen forces during inference. Thanks to the design of the method, we are able to model different garment categories, both tight and loose, with self and body collisions in a topology independent way that does not require an underlying skinned body mesh.

In summary, our main contributions are:
\begin{itemize}
    \item A novel neural architecture with split responsibilities for force modeling and force integration, which allows for integrating physics-based forces in a topology invariant fashion resulting in a method that generalizes to unseen settings.
    \item The method is able to resolve collisions with arbitrary geometries using either triangle mesh or signed distance field (SDF) representations and, for the first time using a neural approach, is able to resolve self collisions between multiple garments.
    \item The resulting simulation pipeline retains the controllable, physics-based, well understood material models, providing the user with meaningful and intuitive control parameters (i.e. material properties).
    \item The integrator module is trained in an unsupervised way, which allows for efficient learning without the need for expensive and hard to obtain training data.
\end{itemize}

\section{Related Work}

We provide an overview of relevant work related to the modeling of physical phenomena using neural networks, cloth detail enhancement and neural methods for generating garment deformations.

\subsection{Neural Networks for Modeling Physical Phenomena}

Neural networks have been successfully used to model granular material and fluids \cite{scarselli2008graph, li2018learning} using particle based approaches and graph neural networks \cite{battaglia2018relational}. Several other methods focus on the modeling of fluid dynamics \cite{thuerey2020deep, um2018liquid}. \cite{pfaff2020learning} introduced a mesh-based method for the simulation of several phenomena using graph neural networks with several follow up works \cite{fortunato2022multiscale, sanchez2020learning, shao2021accurately}.

\subsection{Cloth Geometry Enhancement}

Enhancing details on cloth geometry is a long-standing research problem with early work by \cite{cutler2005art} who presented a procedural wrinkling model capable of adding art-directed wrinkling in a production setting. \cite{bergou2007tracks} proposed a method where constrained Lagrangian mechanics are used to add physically-based details to animated thin shells. \cite{muller2010wrinkle} proposed a simple and fast method to add wrinkles to dynamic meshes by attaching a higher resolution wrinkle mesh to the coarse base mesh. \cite{kavan} presented a method where enhancement is achieved by learning linear upsampling operators for physically-based cloth simulations. \cite{remillard2013embedded} proposed a method to add wrinkling to composite objects consisting of a soft interior and harder skin. Since then, many follow-up works have been presented. \cite{rohmer2010animation} leverage the stretch tensor computed on the coarse animation to add temporally coherent wrinkles. Similarly, \cite{gillette2015real} present a method to add dynamic wrinkling to coarse animated cloth using a two stage stretch tensor estimation process. Cloth details can be enhanced using tension field theory to model coarse geometry after which the amplitude and phase of the fine wrinkling is added \cite{chen2021fine}.
Recently, \cite{wang21gpu} explored specialized techniques for cloth simulation on the GPU using grid-aligned meshes to gain an edge at reconstructing fine wrinkles at submillimeter levels.
Other methods rely on neural networks operating on 3D geometry \cite{NeuralSubdivision, zhang2021deep}. \cite{lahner2018deepwrinkles} present a data-driven approach to enhance detail encoded in a normal map texture. An image-to-image neural network is trained to enhance detail in image space. 

\subsection{Neural Garment Deformations}
Learning-based methods aim at predicting a garment's draping over a given body mesh. Several methods rely on the SMPL~\cite{loper2015smpl} parametric body shape model, along with its rigging and skinning functions. In practice, this means that such methods are able to cast the ML-draping problem as that of predicting corrective garment deformations. These per-vertex deformations are added to the garment's rest-pose vertex position, and are then skinned. Initial methods train a garment prediction model by regressing ground-truth vertex positions, calculated using high-fidelity physics simulation data~\cite{patel2020tailornet}; while relying on fixed skinning weights, transferred from the SMPL weights. The fixed skinning weights limit the garment vertices to move based on their rest pose location. This assumption is alleviated by computing post-deformation skinning weights, utilizing  a prediction network~\cite{santesteban2021self}. Per-pose predictions do not take into account the dynamic nature of garment deformations. Therefore, \cite{santesteban2019learning} utilize a recurrent model whose predictions depend on past poses too. Such models require per-garment training, necessitating multiple ground-truth simulations. \cite{bertiche2021pbns} alleviates this requirement, by training in an unsupervised setup. The loss is cast as a set of physical potentials to minimize: stretching, bending, gravity, and body-cloth collision. These losses are differentiable, and therefore can be back-propagated to optimize the network's weights. \cite{santesteban2022snug} and \cite{bertiche2022neural} both add an inertia loss to address temporal consistency. The input of \cite{santesteban2022snug} further takes not only the parameterized body-pose, but also its shape. However, the network is still specialized per-garment. \cite{de2022drapenet} alleviates this requirement by predicting a latent code for any given garment. It thus generalizes over body shape, body pose, and garment type. Reliance on body-based skinning amounts to limiting the garment to move in correspondence to the underlying body. However, for loose-garment this is not always the case. To tackle this, \cite{pan2022predicting} creates for a garment a new set of joints and corresponding per-vertex skinning weights, based on ground-truth simulations on a variety of motions. Draping prediction then amounts to predicting the joints translation and rotation parameters. The above methods rely on a parametric body representation. This limits their applicability to draping multiple garments layers or stylistic (non-human) avatars. \cite{zhang2022motion} addresses this by representing the underlying body as a set of sampled points, while \cite{d2022n} separately encodes the input body and garment meshes using graph-convolution networks. This separate encoding does not take into account body-garment interactions. To alleviate this shortcoming, \cite{grigorev2022hood} adds body-garment graph edges, and uses hierarchical message-passing. Body-garment collisions can also be solved by learning a collision-free generative deformation space~\cite{santesteban2021self}. ULNeF~\cite{santesteban2022ulnef} generalizes to multiple garment by predicting corrective terms to the garments' implicit representation; but is limited to running on human shapes in canonical pose. 

The above ML-based garment draping methods achieve impressive results. However, they are all limited in generalizing to arbitrary-posed bodies with arbitrary layers of interwoven garments and clothing items (such as a tucked-in shirt, layered with a suit, tie and a pocket handkerchief as in Fig.~\ref{fig:teaser}). Our method allows, for the first time, to achieve realistic draping of complex topologies, in a self-supervised manner (see full comparison in Table~\ref{tab:comparison}).

\section{Method}
To evaluate \emph{PhysGraph}, we focus on quasi-static simulation of cloth using elastic energy and contact penalty potential minimization. 
Given an energy potential, a classical method would 
iterate over configurations following the negative energy gradients (forces) until it finds the optimal point.
Here, we decouple the \emph{force module} responsible for directly differentiating energy potentials, and the \emph{integration module}, which integrates the resulting forces into a displacement vector for all vertex positions.
Given these complementary responsibilities of the modules, our method provides great flexibility to the system being modeled.
The method is agnostic to both the type of physical forces modeled by the force module and, the mesh connectivity at inference time, allowing it to generalize to different types of potentials even after the network is trained.
The central building block of our architecture is a recurrent mesh graph network. We initialize the integration module with an upsampled version of the physical simulation given a coarse mesh. This way, the large scale dynamic behavior of the system is captured in the coarse mesh while our architecture creates finer scale details that are governed by the static equilibrium of forces. These coarse meshes can be obtained using classical mesh based simulation or other methods like artist models and linear blend skinning output.
 
Our key contribution is an algorithm that performs several iterations of force computation and integration to find an approximate minimizer of the potential energy provided by the force module during inference. To this end, the integrator leverages a graph network architecture \cite{pfaff2020learning} which we use in a recurrent fashion. The subsequent sections will introduce the force module, the integration module and the graph network architecture.

\subsection{Force module} 
\label{sec:force}
We credit the generalizability of our method to the separation of the force formulation and integration. The force generation module outputs forces based on user-specified potentials given the current nodal configuration where forces are accumulated at the nodal level, enabling parallelization of the force computation. The conservative force potential $\Phi$, is responsible for the forces $\textbf{F}=-\nabla_\textbf{X}\Phi$ acting on the system. Different potentials can be used to model different physical phenomena.

\subsection{Mesh-Based Graph Networks}
\label{sec:gn}
\begin{figure}
\includegraphics[width=0.95\linewidth]{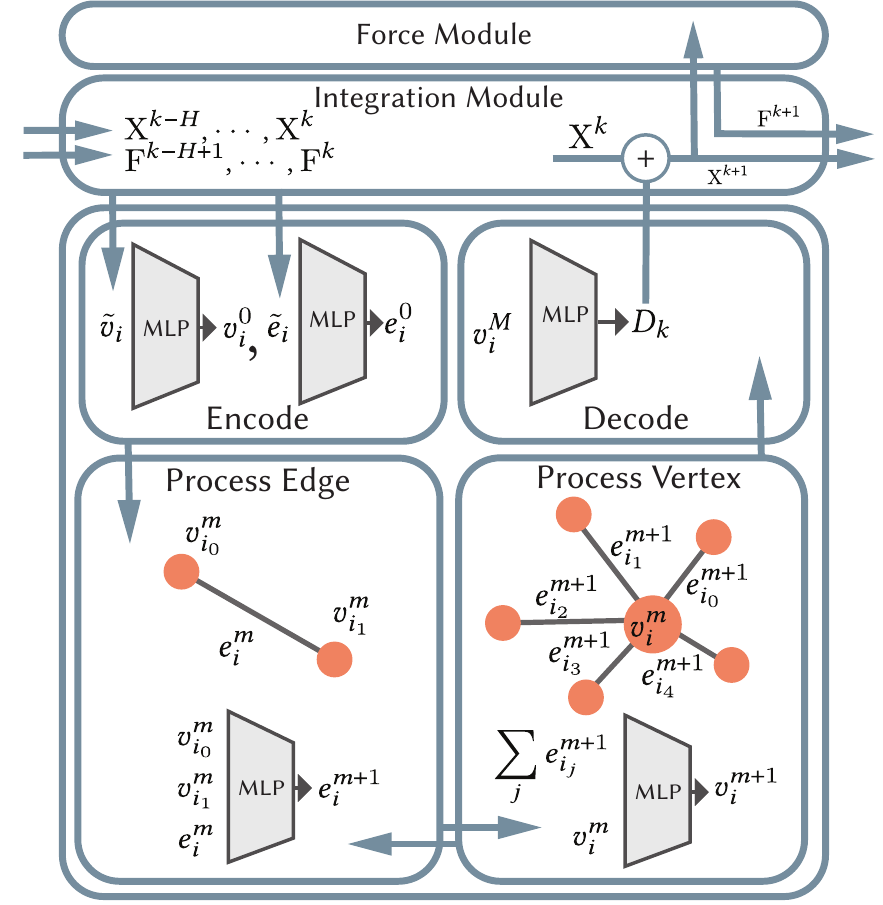}
\caption{Our approach consists of three components. The force module (Section \ref{sec:force}) which evaluates per-node forces for a mesh configuration. The integration module (Section \ref{sec:integration}) constructs initial feature vectors $\tilde{v}_i$ and $\tilde{e_i}$ based on force and configuration information from previous iterations, passes these quantities to a graph neural network (Section \ref{sec:gn}) and retrieves new displacements $D_k$. The graph network iterates between an edge processing step that distributes vertex information to edges and an edge processing step which distributes edge information to adjacent vertices. These distribution steps are iterated $M$ times. The integration module iterates $K$ times and finally returns an estimate for the quasi-static equilibrium state $\mathbf X^K$.}
\label{fig:overview}
\end{figure}
The simulation mesh can be interpreted as a graph $G=(V,E)$ onto which we encode vertex and edge features consisting of 128 values each. The graph network operates in three phases: Encode, Process and Decode.
%
\paragraph{Encode} Vectors of per edge $\tilde{e}_i$ and per vertex features $\tilde{v}_i$ form the input to the graph network. By applying two multilayer perceptrons (MLP), one to each vertex feature and one to each edge feature, we obtain initial features $v_i^0$ and $e_i^0$.
%
\paragraph{Process}
We use a fixed amount of $M=10$ message passing iterations. Each iteration $j$ consists of two steps computing new edge and vertex features. First, $e_i^j$ is computed by passing information from vertices to edges. In the second step, information is passed from edges to adjacent vertices to build $v_i^j$, see Figure \ref{fig:overview}. Each message passing iteration uses two MLPs with weights that are shared between all vertices and edges, respectively.
%
\paragraph{Decode}
A final MLP is used to decode the final vertex features $v_k$ into the final displacement vectors.
\newline

The set of learnable parameters $\theta$ are the weights of the five MLPs used during the three phases. 
\subsection{Integration Module}
\label{sec:integration}
The physical configuration of the mesh is given by its embedding which can be represented as a matrix of stacked position vectors $\mathbf X \in \mathbb R^{n \times 3}$ where $n$ is the number of vertices in the mesh and $\mathbf X_i$ represents the position of the $i-$th vertex. We assume that the mesh has a rest configuration $\overline{\mathbf X} \in \mathbb R^{n \times 3}$, which by definition, experiences no internal forces.
%
The goal of the integration module is to find an approximation to the static equilibrium configuration $\mathbf X^* = \operatorname{argmin}_{\mathbf X} \Phi(\mathbf X)$ with potential energy $\Phi$. Starting with initial vertex positions $\mathbf X^k$ with $k=0$, we produce new positions $\mathbf X^{k + 1}$ by using the pre-trained graph network. Throughout the simulation, the graph network has access to the mesh connectivity corresponding to the set of graph edges $E$. For each configuration, we can query the force module to obtain corresponding forces $\mathbf F^k = -\nabla_\X \Phi ( \X^k)$. 
The inputs to the graph network are per vertex and edge features.
For each vertex $i$ we construct the feature vector $\tilde{v}_i$ by concatenating first order differences of the last $H$ configurations and the corresponding force vectors
\begin{equation} 
\begin{split}
    \tilde{v}_i = ( \mathbf X^{k-1}_i - \mathbf X^{k}_i, \cdots,  \mathbf X^{k-H}_i - \mathbf X^{k-H+1}_i,\\ \mathbf F^k_i, \cdots,\mathbf F^{k - H + 1}_i ) .
\end{split}
\end{equation}
The edge features $\tilde{e}_i$ for the edge connecting vertex $r$ and $s$ are comprised of position differences for the current configuration and the rest state as well as their lengths 
\begin{align}
    \tilde{e}_i = \begin{pmatrix}\mathbf X^k_r - \mathbf X^k_s, \|\mathbf X^k_r - \mathbf X^k_s \|,  \overline{\mathbf X}_r - \overline{\mathbf X}_s, \|\overline{\mathbf X}_r - \overline{\mathbf X}_s \|\end{pmatrix},
\end{align}
where $\|\cdot\|$ is the Euclidean norm.
The network outputs displacements $\mathbf D^k$ that define the next state via $\mathbf X^{k+1} = \mathbf X^{k} + \mathbf D^{k}$.
After $K=5$ iterations we obtain an approximation of the equilibrium state $\mathbf X^K$. 

\subsection{Training phase}

We train the integration module with a dataset of small physical systems. These act as an input to the recurrent model, consisting of $K$ recurrent force calculation and integration blocks sharing the same network parameters of the trainable integration module. The training is supervised by requiring the minimization of the potential, summed over all the intermediate states. 





\section{Garment Detail Enhancement}


We apply PhysGraph to the cloth upsampling problem. A given coarse resolution cloth mesh is first subdivided. Then nodal forces are computed using the force module and integrated into nodal displacements by the integration module. This is repeated iteratively over multiple steps, which allows local forces to propagate throughout the rest of the mesh.
We show that the method remains effective regardless of whether the coarse mesh is obtained through low resolution cloth simulation or other methods such as artist animation or through procedural or skinning approaches. In this section, we define the potentials used to model fabric elasticity and contact. 

\subsection{Garment Potentials}
To demonstrate cloth modelling, we use springs to model stretching, dihedral angle penalty to model
\begin{wrapfigure}[5]{r}[20pt]{0.2\textwidth}
\vspace{-12pt}
\hspace{-40pt}
\centering
\includegraphics[width=0.2\textwidth]{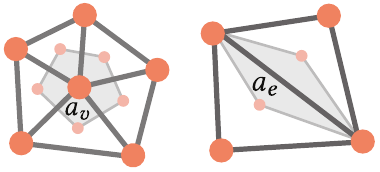}
\end{wrapfigure}%
bending and penetration penalty to model contact, although the force module can be any material model
that produces forces.
The potentials are weighted based on element
area, which we define to
be the area of a barycentric subdivision (light gray areas in the inset). This way the total area of all edge $a_e$ and vertex $a_v$ weights, respectively, sum up to the total surface area $A$.

\textbf{Stretching} is modelled using edge aligned springs with net elastic potential
\begin{equation}
   \Phi_{s} = \frac{k_s}{2A} \sum_{e \in E} a_e (l(e) - l_0(e))^2,
\end{equation}
where $k_s$ is the spring stiffness, $l(e) = \|\X_i - \X_j\|$ is the edge length for an edge $e=(i, j)$  and $l_0(e) = \|\overline \X_i - \overline \X_j\|$ is its rest-length.
The \textbf{bending} potential is defined by the dihedral angles $\theta_d$ formed between the normal vectors to the triangles in each dihedral element
\begin{equation}
        \Phi_b = \frac{k_{b}}{2A} \sum_{e \in D} a_e \theta_d^2
\end{equation}
with bending stiffness $k_b$ where $D$ is the set of interior edges corresponding to dihedral elements.
The \textbf{gravitational} potential is defined by
\begin{equation}
        \Phi_g = - \frac{g}{A} \sum_{v \in V} a_v m_v z_v, \quad \text{where } m_v = \rho a_v,
\end{equation}
and $z_v$ is the coordinate along the gravity axis and $\rho$ the mass density.
The \textbf{external contact} potential is modeled using the signed-distance-function $SDF(\cdot)$, which measures the signed distance (negative inside, positive outside) to the surface of a set of colliders in the system. The contact potential $\phi_v = -\operatorname*{min}(SDF(\X_v), 0)$ is accumulated over all potentially violating vertices $v$ with 
\begin{equation}
\begin{split}
    \Phi_{ec} &= \frac{k_{ec}}{A_{ec}}\sum_{v \in V} a_v \phi_v,\quad A_{ec} = \sum_{v\in V}  a_v  (1 - \ind(\phi_v)),
\end{split}
\end{equation}
and $k_{ec}$ is the external contact penalty stiffness.
The zero-set indicator function $\ind$ evaluates to $1$ for $0$ and to $0$ otherwise.
Finally, the \textbf{self collision} potential is modeled by radial compression springs with rest-length $R$ around each vertex. A compression spring between vertices $u,v \in V$, modelled by the potential $\psi_{u,v} = \max(R - \|\X_u - \X_v\|, 0)$, exerts a force in the outward radial direction when compressed, which happens when distinct vertices become closer than $R$ apart. The total energy is defined by
\begin{equation}
%
    \begin{split}
    \Phi_{sc} = \frac{k_{sc}}{A_{sc}}\!\!\!\! \sum_{\substack{u,v \in V \\ u \notin \mathcal N_d(v)}}\!\!\!\! (a_v + a_u)  \psi_{u,v}^2, \\
    A_{sc} = \!\!\!\!\! \sum_{\substack{u,v \in V \\ u \notin \mathcal N_d(v)}}\!\!\!\! (a_v + a_u) (1 - \ind( \psi_{u,v}))
    \end{split}
\end{equation}

where $k_{sc}$ is the self-collision penalty stiffness, and we consider only interactions of vertices which are not neighbors on the mesh within some d-ring of $v$ denoted $\mathcal N_d(v)$.

\subsection{Training}

The method is trained using patches sampled from a dynamically simulated t-shirt on a moving human body, example patches are shown in Figure~\ref{fig:teaser}. The patches are subdivided using a self-similarity subdivision scheme, increasing the mesh resolution by a factor of 16 and, by linearly interpolating the coordinates. We use $K=5$ recurrent blocks in our experiments. We stress that we only include forces resulting from the stretch and bending potentials $\Phi=\Phi_s+\Phi_b$ at training time. To account for the fact that the patch is a sub-system of the larger full-cloth system and prevent the flattening of the patches in the absence of the rest of the cloth-system, we used fixed boundary conditions while training. 




\section{Results}

Due to the design of our method, we are able to train our integrator network using only the internal cloth potentials and their resulting forces. At inference time, the trained network is capable of ingesting numerous forces from a variety of sources to produce plausible results. We showcase this by applying our trained integration module to a variety of novel forces. We show results with forces such as gravity, body collision and self collision. We demonstrate the effectiveness of PhysGraph for several variations of the cloth enhancement application with a complete multi-layered garment example shown in Figure~\ref{fig:dress_different_forces} and the supplemental material. All results were generated using a single trained network which was only exposed to elastic forces at training time. Note that the method is not limited to these specific examples.

\subsection{Coarse Simulation Enhancement}

\begin{figure}
    \centering 
    \includegraphics[width=0.4\textwidth]{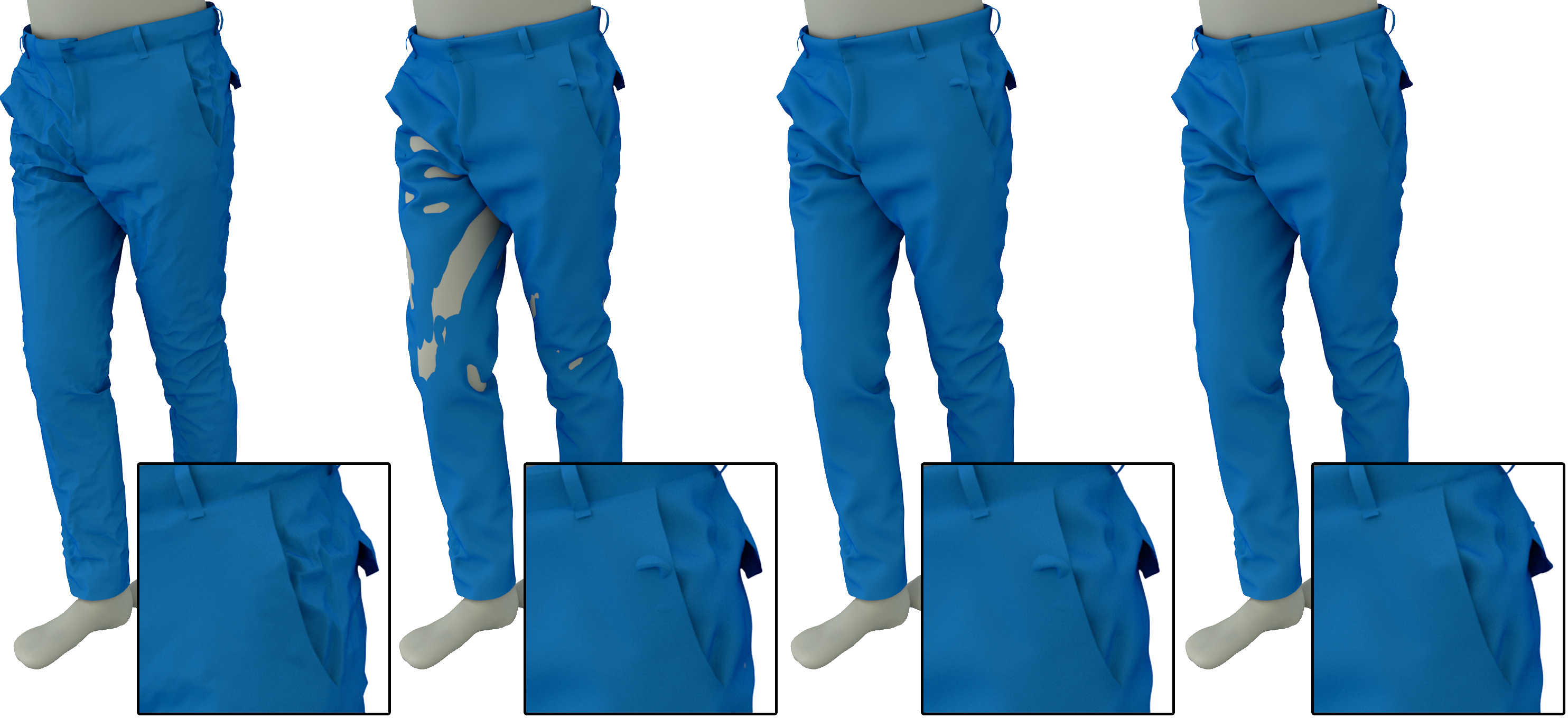}
    \caption{From left to right: coarsely simulated input, inferring with stretch and bending forces only, stretch and bending and body collisions and finally, including all forces on the right. The model was trained with elastic energy forces only and generalizes to include collisions.} 
    \label{fig:pants_different_forces}
\end{figure}

\begin{figure*}
    \centering 
    \includegraphics[width=0.8\linewidth]{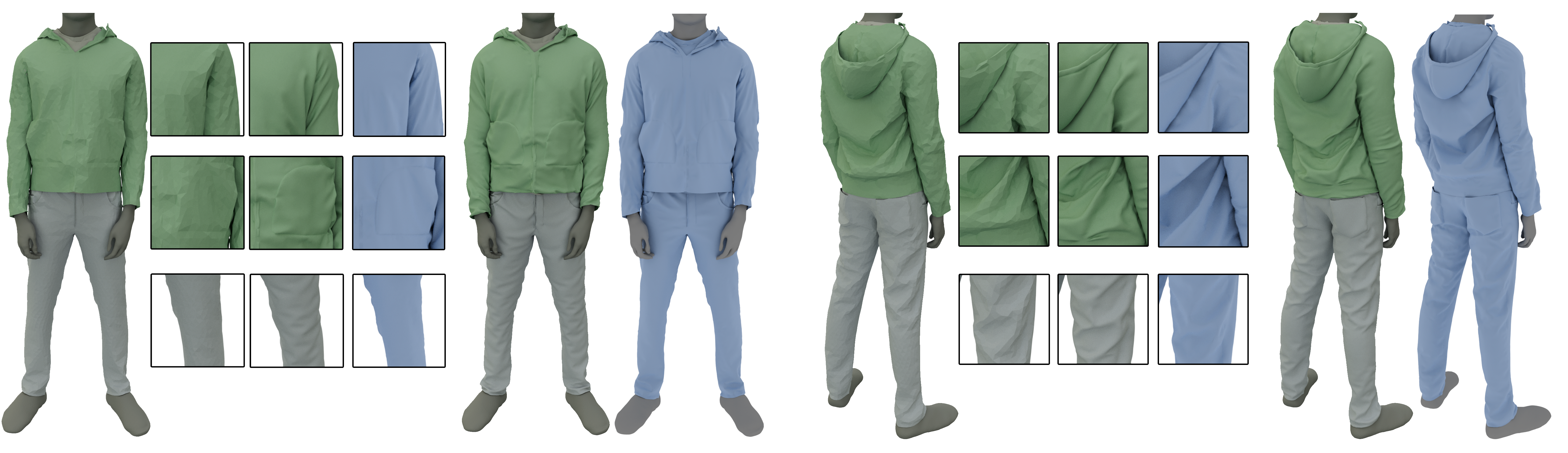}
    \caption{In contrast to methods that rely on skinning based techniques, our method naturally handles loose clothing such as the hood of a sweater. We show a front and back view, before (left) and after (middle) upsampling as well as ground truth simulation in blue. Note that the clothing is not expected to be exactly the same due to different material parameter selection.}
    \label{fig:dress_different_forces}
\end{figure*}

We show results for the cloth geometry enhancement in Figure~\ref{fig:pants_different_forces} which shows the resulting geometries after integrating different force potentials ranging from simply including elastic energy potentials, which were included at training time, to a full model with body and self-collisions, which includes several forces not seen during training. Our method is garment independent and works for loose clothing as can be seen in Figure~\ref{fig:dress_different_forces}.

\subsection{Linear Blend Skinning Enhancement}

\begin{figure}
  \centering
    \includegraphics[width=0.9\linewidth]{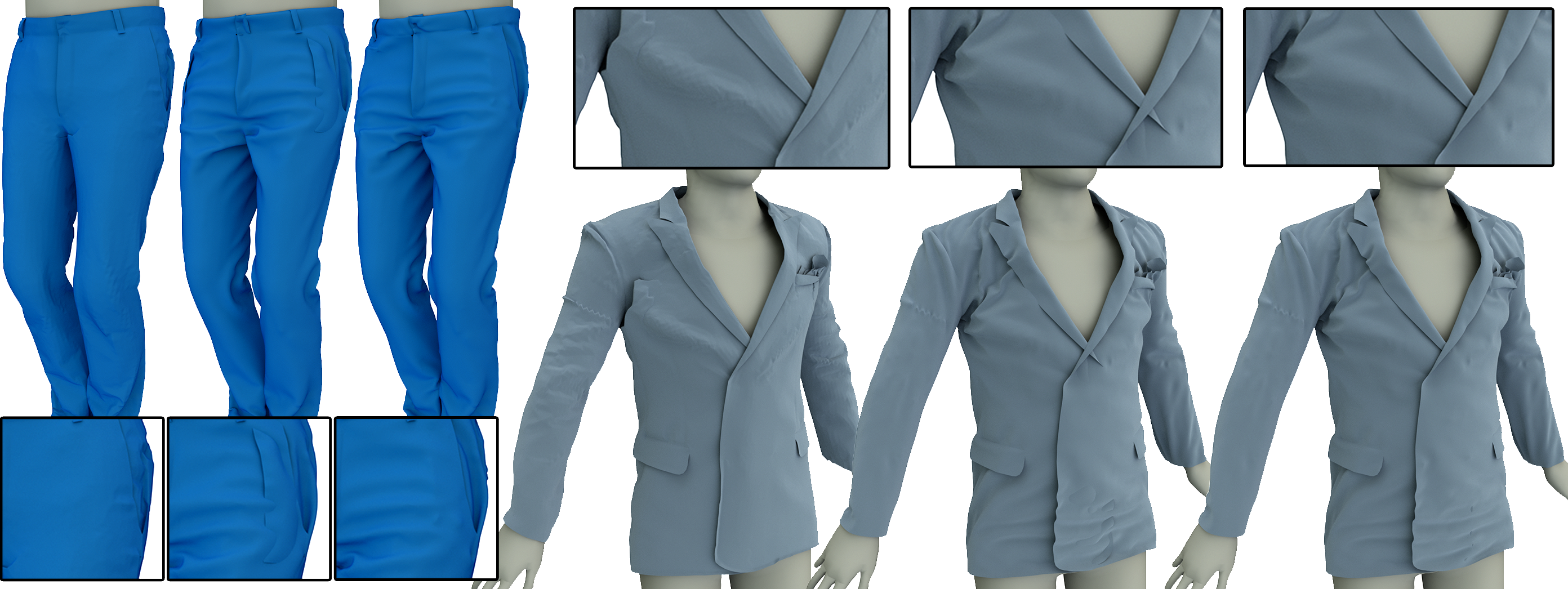}
    \caption{We demonstrate that our method is capable of enhancing geometry obtained from linear blend skinning (left), the middle shows the enhanced mesh without self collisions and the rightmost shows the predicted garment incorporating all forces. Note the realistic wrinkling added by our method while preserving the overall shape and remaining collision free. The enhancement model includes both body and self collision forces. Note that for the jacket, the method is even capable of removing the skinning artifacts near the armpit.}
    \label{fig:linearBlendSkinning}
\end{figure}

To demonstrate generalization to the model input, we show that the coarse geometry does not need to be obtained from a simulation and lower cost methods can be used. We showcase the efficacy of the method on garments posed using linear blend skinning which is known to have several issues which distorts the mesh in non-physical ways. Nevertheless, Figure~\ref{fig:linearBlendSkinning} shows that our method is capable of producing visually pleasing high resolution meshes where self-collisions are resolved.

\subsection{Material Generalization}

 \begin{figure}
    \centering 
    \includegraphics[width=0.4\textwidth]{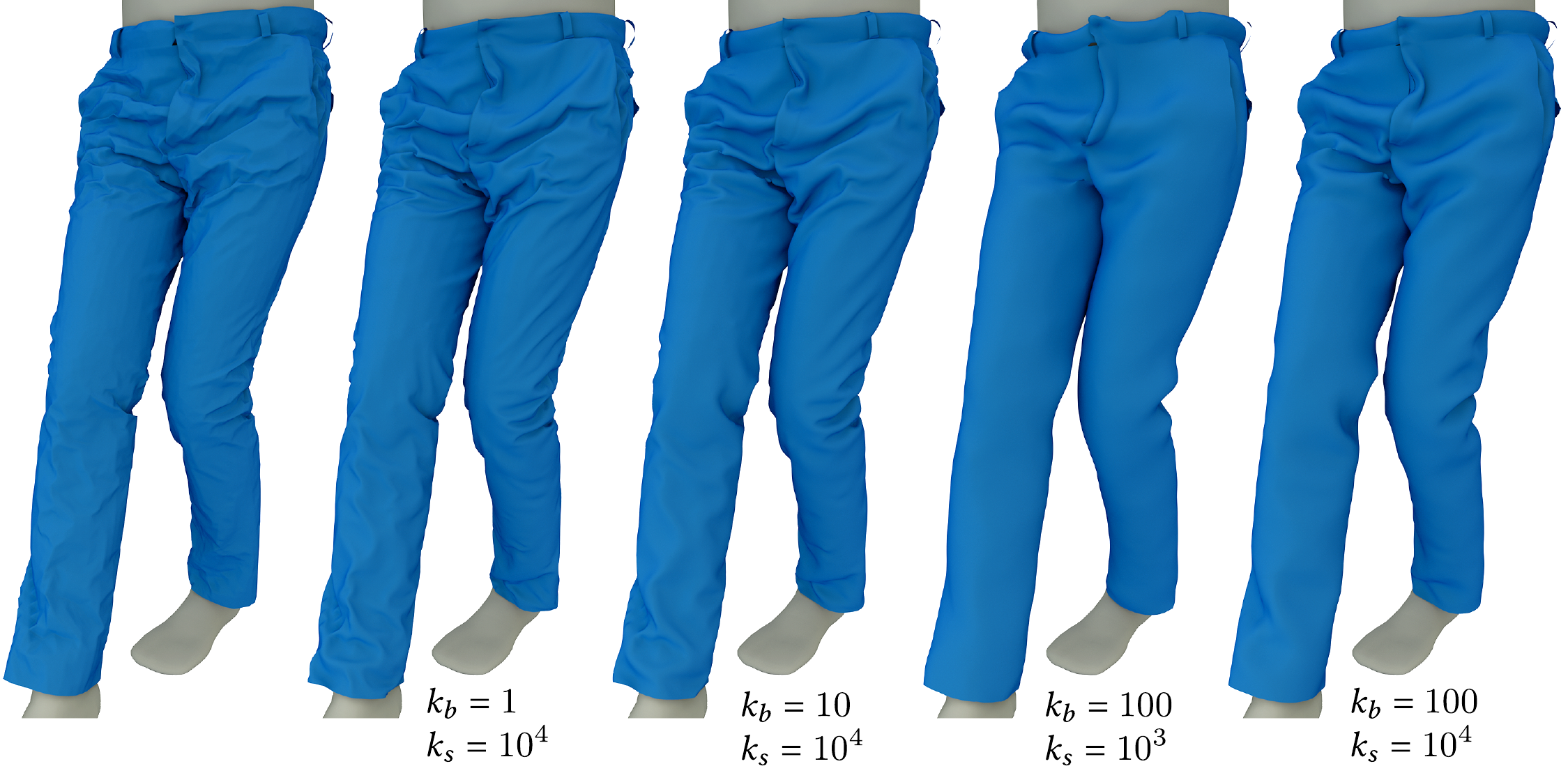}
    \caption{We demonstrate the ability of enhancing garment geometry using different materials at inference time. This example shows the apparent visual difference of varying bend and stretch stiffnesses allowing the user to generate a potentially expensive coarse sequence once and adjust materials afterwards. Our method produces plausible results where higher bending stiffness correctly corresponds to bigger folds.}  
    \label{fig:pants_change_material}
\end{figure}

Due to the design of the separate force module, we are able to modify the material properties at inference time. Figure~\ref{fig:pants_change_material} shows different material settings for a pair of pants which are all generated from the same coarse input geometry. Although our training set uses the same model with a one set of parameters for all training samples, our method can work with other properties, because the integration module does not depend on the origin of the forces.

\subsection{Collision Geometry Generalization}

 \begin{figure}
    \centering
    \includegraphics[width=0.4\textwidth]{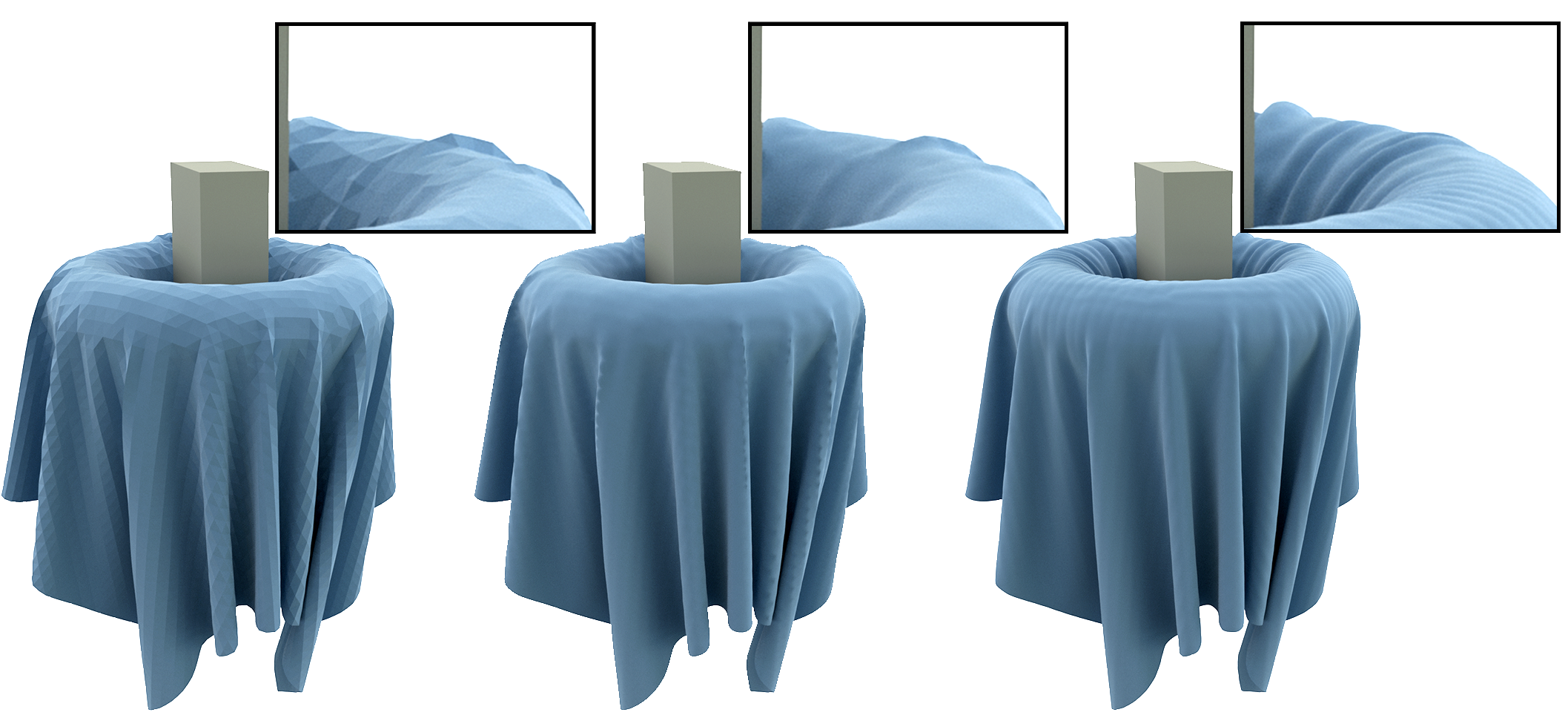}
    \caption{We show that our method generalizes to different collision geometries during inference. From left to right, we show the original low resolution input, a mesh obtained using Loop subdivision~\shortcite{loop1987smooth}, and our method.
    Note how our method adds detail compared to a subdivision which simply smooths. Our method maintains the overall shape of the low resolution but adds detail, preserving artistic intent.
    } 
    \label{fig:genericCollider}
\end{figure}

Our method generalizes to different collision geometries as shown in Figure~\ref{fig:genericCollider}. We show rich cloth interaction with a rigid block, pushing the cloth through the center of a torus. These colliding geometries are completely novel with respect to those at training time, yet, the method produces correct results.

\subsection{Self Collisions}

Interactions of fabric with itself are ubiquitous for cloth simulations as garments are often layered and display complex interactions. Therefore, it is essential to model them appropriately. We demonstrate the importance of including self collisions in Figures~\ref{fig:teaser},~\ref{fig:pants_different_forces}~and the supplemental material, where we show that our model is capable of including these forces, providing clean, intersection free meshes as shown on the right.

\subsection{Convergence Analysis and Performance}\label{sec:convergence_analysis}

\begin{figure}
  \centering
   \resizebox{0.7\columnwidth}{!}{
     \centering
    \vspace{-6pt}
    \hspace{-0.9cm}
\begin{tikzpicture}
	\begin{axis}[width =\linewidth,
	height = 0.7\linewidth,
    legend columns = 3,
    legend cell align=left,
    legend style={font=\footnotesize, at={(1.0, 0.98)}, anchor=north east, line width=0.4pt, draw=.!60!white},
   	xlabel= iteration,
        ylabel= potential,
        ylabel style={yshift=-0.5cm, font=\small},
        xlabel style={yshift=0.2cm, font=\small},
        axis line style = {thick, color=.!60!white},
		axis y line=left,
	axis x line=bottom,
        ymajorgrids = true,
        tick align=inside,
	tick label style = {font=\small},
        ytick = {0.2, 0.4, 0.6, 0.8, 1.0, 1.2},
        yticklabels = {0.2, 0.4, 0.6, 0.8, 1.0, 1.2},
		ymin = 0.0,
		ymax = 1.4],
	\addplot[ultra thick, color=col1, mark=*, mark size=0.5] table [x index=0, y index=1,col sep=tab] {data/plotData};
 \addlegendentry{adam $10^{-2}$}
	\addplot[ultra thick, color=col2,mark=*,   mark size=0.5] table [x index=0, y index=2,col sep=tab] {data/plotData};
 \addlegendentry{adam $10^{-3}$}
	\addplot[ultra thick, color=col3, mark=*, mark size=0.5] table [x index=0, y index=3,col sep=tab] {data/plotData};
 \addlegendentry{adam $10^{-4}$}
 	\addplot[ultra thick, color=col4,mark=*, mark size=0.5] table [x index=0, y index=4,col sep=tab] {data/plotData};
  \addlegendentry{gd $10^{-1}$}
	\addplot[ultra thick, color=col5,mark=*,   mark size=0.5] table [x index=0, y index=5,col sep=tab] {data/plotData};
 \addlegendentry{gd $10^{0}$}
	\addplot[ultra thick, color=col6, mark=*, mark size=0.5] table [x index=0, y index=6,col sep=tab] {data/plotData};
 \addlegendentry{gd $10^{1}$}
 	\addplot[ultra thick, color=col7, mark=*, mark size=0.5] table [x index=0, y index=7,col sep=tab] {data/plotData};
  \addlegendentry{PhysGraph}
	\end{axis}
\end{tikzpicture}
   }
    \vspace{-8pt}
    \caption{We demonstrate that our method converges faster and to a lower potential than several gradient-descent optimizers, with varying base learning rates. For each baseline optimizer, we show increased learning rates until they diverge.}
    \label{fig:convergenceComprison}
\end{figure}
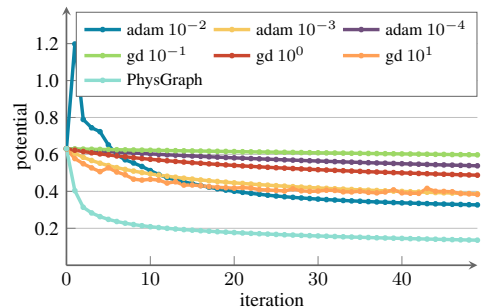

We highlight the effectiveness of PhysGraph by comparing convergence with respect to two baseline optimizers: Adam and gradient decent. We showcase various learning rates in Figure~\ref{fig:convergenceComprison} for which they still converge. Note how our learned integrator is the most effective. While we don't claim performance improvements compared to state-of-the-art, we show competitive timings in Table~\ref{tab:performance} for the different modules of our method for a variety of resolutions.

\subsection{Comparisons To Related Work}
\begin{figure}[h]
    \centering 
    \includegraphics[width=\linewidth]{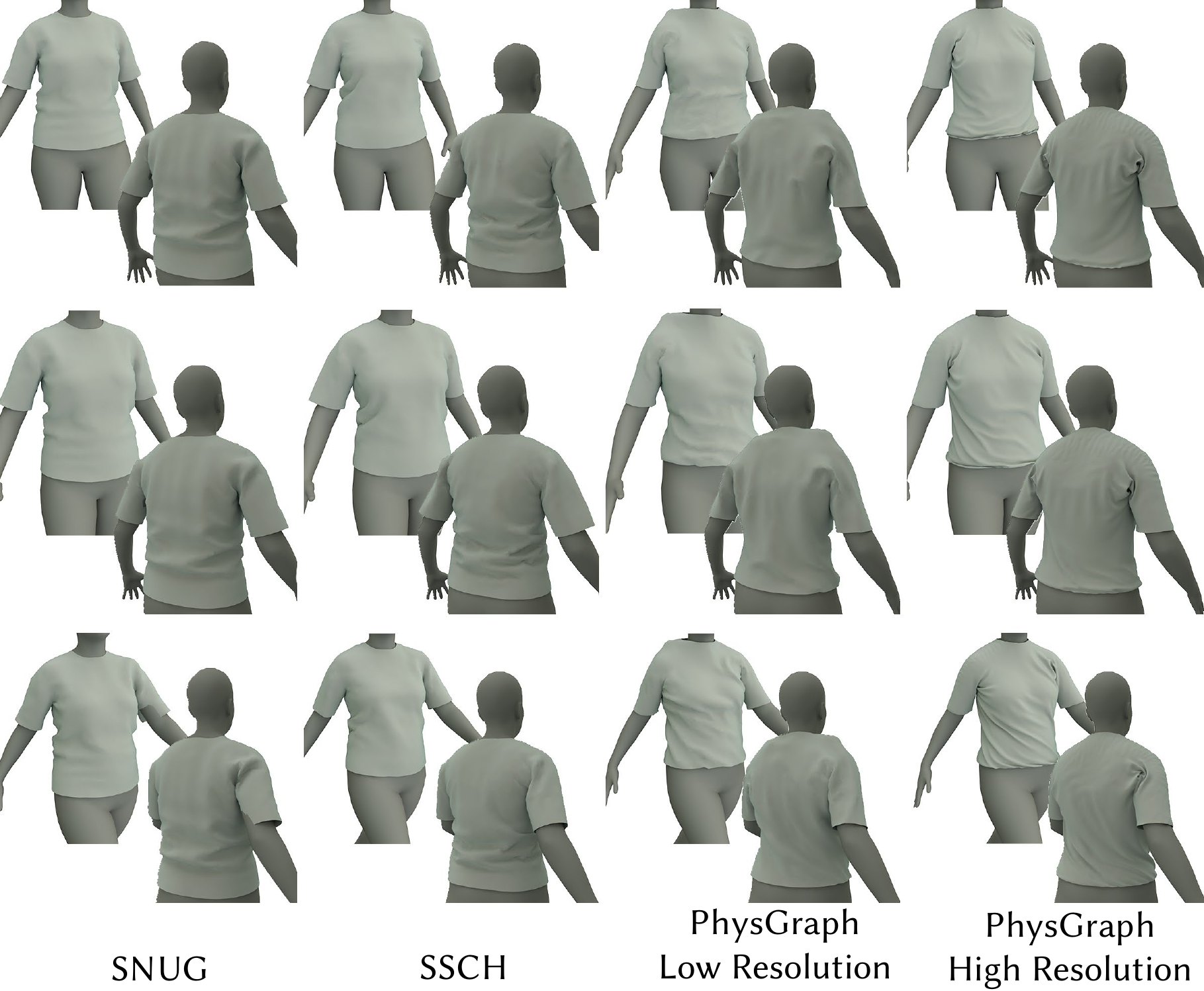}
    \caption{Comparison to skinning-based ML-draping methods of three different frames of the CMU-07-02-poses sequence from AMASS~\cite{mahmood2019amass}. From left to right:   SNUG~\cite{santesteban2022snug}; SSCH~\cite{santesteban2021self}; PhysGraph low resolution prediction; PhysGraph high resolution refinement. Note how SSCH and SNUG show similar wrinkling regardless of the pose, whereas PhysGraph is capable of adding fine detail in a more physically plausible way at several resolutions where detail varies with pose. Also note that SNUG results in intersections with the body as seen in the bottom row.} 
    \label{fig:compare_to_skinning_baselines}
\end{figure}

We compare PhysGraph (ours) to two recent ML-approaches for garment draping. Both SSCH~\cite{santesteban2021self} and SNUG~\cite{santesteban2022snug} are skinning-based approaches, trained with ground-truth simulations and self-supervision, respectively.  
We use their publicly released t-shirt models for comparison, and emphasize that both models were trained \emph{for this specific garment}. In comparison, our method has not seen this garment during training. Our method takes the skinned mesh as input, and predicts a refined draping. Figure~\ref{fig:compare_to_skinning_baselines} shows a qualitative comparison and Table~\ref{tab:quantitative_comparison} provides quantitative comparison of the system potentials for each method. While both SNUG and SSCH are limited to a fixed topology and resolution since both models are specialized for this specific garment, we are still able to generate detail at several resolutions without having trained on this topology. Furthermore, our method provides qualitatively and quantitatively better results with a more general method which has not been optimized for this particular setting.
Table~\ref{tab:comparison} provides a functional comparison which additionally includes ULNeF~~\cite{santesteban2022ulnef}, Hood~\cite{grigorev2022hood} and MeshGraphNet~(MGN)~\cite{pfaff2020learning}. 

\begin{table}[h]
    \centering
        \resizebox{0.9\columnwidth}{!}{
\begin{tabular}{lcc}
\toprule
&
Total Potential [erg] $\downarrow$ &
Body Collision (\%) $\downarrow$ \\ \midrule
Skinning & 243.2094 & 4.5088 \\
SSCH & 106.2101 & 0.8257 \\
SNUG & 91.3766 & 0.7818 \\
\textbf{Ours} & \textbf{74.4635} & \textbf{0.4813} \\ \bottomrule
\end{tabular}
}
       \caption{\emph{Quantitative comparison with state-of-the-art methods}. $\downarrow$ means a lower value is better. We report the potential of the cloth in [erg] units, using a mass-spring model under gravity, using $k_s=1e4$ erg/cm$^2$, $k_b=10$ erg, and $\rho=0.0187$ gr/cm$^2$. To obtain the rest edge lengths, we use SNUG's rest mesh. The collisions with the body are reported as the percentage of cloth vertices admitting negative values when used as a query to the body SDF, which is defined with respect to the body with a collision margin of 2 mm, similar to SNUG. All the compared potentials are calculated over the same mesh topology taken from SNUG's shirt.}
    \label{tab:quantitative_comparison}
\end{table}

\begin{table}[h]
    \centering

\resizebox{\columnwidth}{!}{
\begin{tabular}{lcccccc}
\toprule
& SSCH & SNUG & ULNeF & MGN & Hood & \textbf{PhysGraph} \\
\midrule
Topology Invariant       & \xmark & \xmark & \cmark & \cmark & \cmark & \cmark \\
Pose Invariant           & \cmark & \cmark & \xmark & \cmark & \cmark & \cmark \\
Force Agnostic           & \xmark & \xmark & \xmark & \xmark & \xmark & \cmark \\
Body-Garment Collisions  & \cmark & \xmark & \xmark & \cmark & \cmark & \cmark \\
Garment self Collisions  & \xmark & \xmark & \cmark & \cmark & \cmark & \cmark \\
Multi Garment Collisions & \xmark & \xmark & \cmark & \xmark & \xmark & \cmark \\
Unsupervised             & \xmark & \cmark & \xmark & \xmark & \cmark & \cmark \\ \bottomrule
\end{tabular}
}
    \caption{\emph{Summary of Related Work.} Our work achieves \emph{all} desirable features.}
    \label{tab:comparison}
\end{table}

\begin{table}[h]
    \centering
    \resizebox{0.9\columnwidth}{!}{
\begin{tabular}{c | cc }
\toprule
Simulated Edges &
Force Module [ms] &
Integration Module [ms] \\ \midrule 
1e3 & 0.036864 & 7.13121 \\ 
1e4 & 0.041984 & 9.744 \\ 
1e5 & 0.051232 & 37.74 \\ 
1e6 & 0.077824 & 343.08 \\ 
\bottomrule
\end{tabular}
}
       \caption{Performance measurements for a varying number of springs in the physical system measured on a NVIDIA RTX A6000 GPU using a CUDA implementation. Timings are reported in milliseconds.}
    \label{tab:performance}
\end{table}

\section{Discussion, Limitations, And Future Work}

We propose a novel method for the integration of physics-based forces using a graph neural network. We demonstrate detail enhancement of coarse cloth geometry which can be obtained from several sources such as simulation, content creation tools or linear blend skinning among others. Our method is capable of modeling garment interactions with itself and other collision objects and we are the first to support collisions with multiple garments simultaneously using their geometry directly without needing an SDF or other representation which introduce several limitations for modeling layered clothing. Although we already demonstrate competitive computation times, we believe that given additional engineering resources, the method has the potential to run even faster since there are active research efforts on improving message-passing architectures efficiency \cite{rahman2021fusedmm, xie2022graphiler} which have already demonstrated the ability to accelerate the computation by two orders of magnitudes.
Like most other methods, our current model is unable to resolve pre-existing self-intersections. However, future work could include untangling forces \cite{baraff2003untangling} as part of the force module.
In the supplemental material, we show how our method performs on a simulated T-shirt sequence, where our generated detail is already mostly temporally coherent with the exception of areas with clustered collisions. This is encouraging given that the method operates per frame independently without exploiting temporal information. In the future, we aim to resolve these remaining issues by accounting for temporal features during training and inference. 
In the future, we want to apply our method to physical phenomena beyond garment simulation such as those potentials required to model hair and volumetric materials.




{\small
\bibliographystyle{ieee_fullname}
\bibliography{egbib}
}

\clearpage
\setcounter{page}{1}


\twocolumn[{%
\renewcommand\twocolumn[1][]{#1}%
\begin{center}
    \centering
    \captionsetup{type=figure}
    \includegraphics[width=0.9\linewidth]{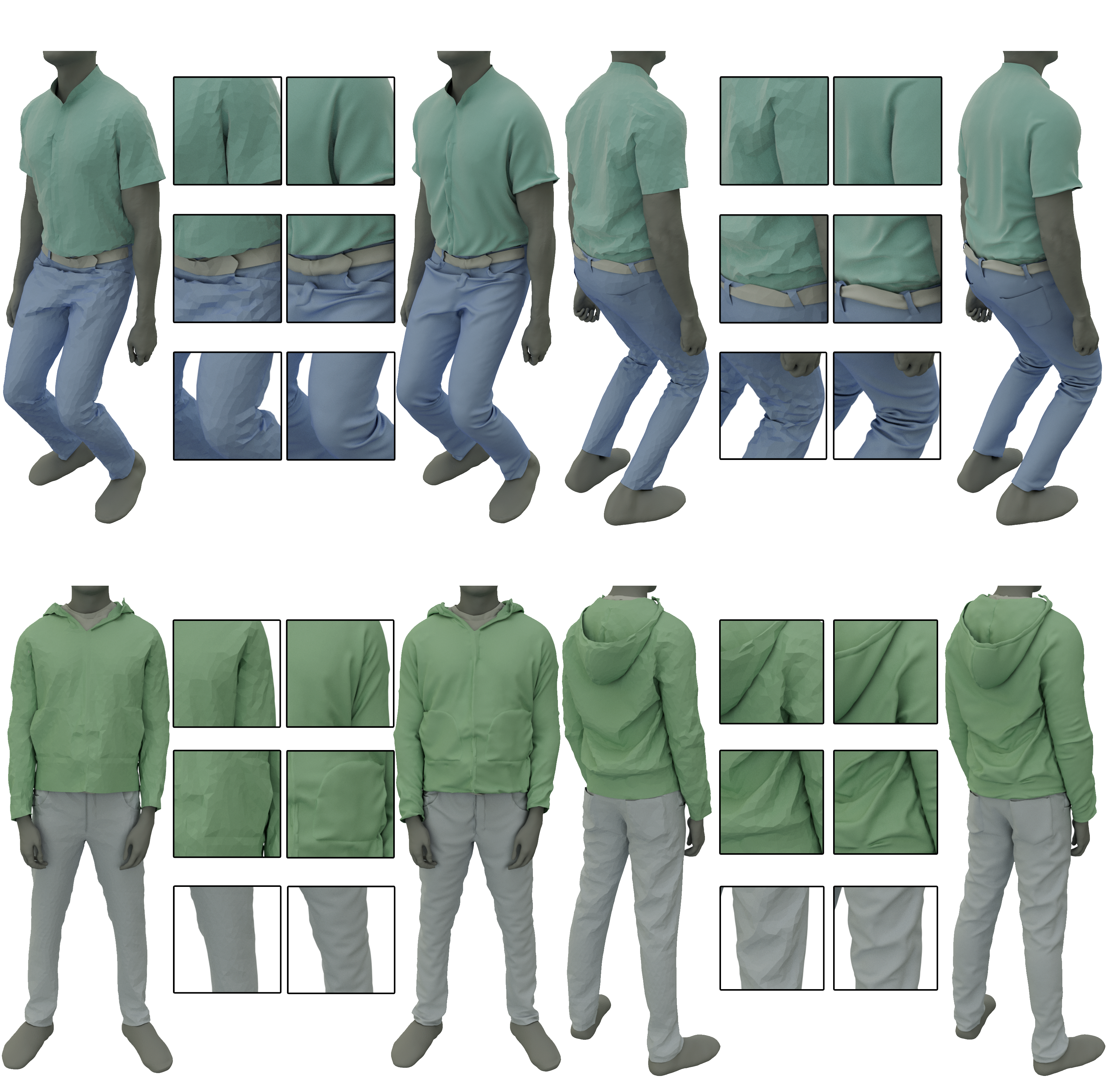}
     \captionof{figure}{We show an additional example of the cloth enhancement process using PhysGraph to demonstrate that our method scales to complicated multi-layer outfits. We demonstrate a tucked in shirt with belt and pants. Note how our model is capable of resolving collisions with small geometric features such as the belt loops and pockets.}
    \label{fig:fullPageResult}
\end{center}%
}]

\twocolumn[{%
\renewcommand\twocolumn[1][]{#1}%
\begin{center}
    \centering
    \captionsetup{type=figure}
    \includegraphics[width=0.6\linewidth]{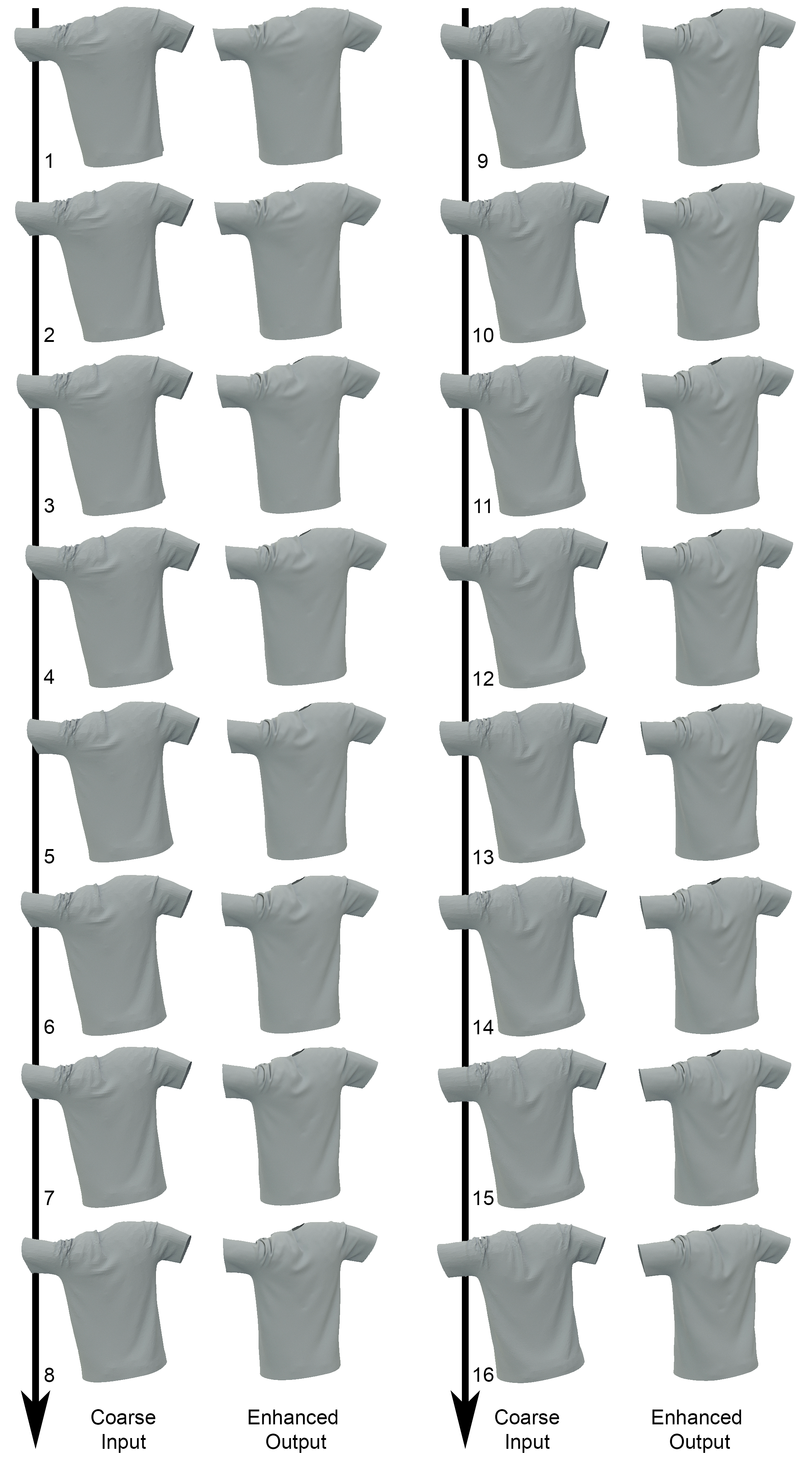}
    \captionof{figure}{We show several consecutive frames of an enhanced cloth sequence. Despite operating per frame, our method shows mostly temporally consistent results with the exception of collision heavy regions. Please refer to the supplemental video for the full result.}
    \label{fig:tshirtSequence}
\end{center}%
}]

\end{document}